\documentclass[a4paper,11pt]{article}
\usepackage{pos}
\usepackage{floatrow}  
\newcommand{\hc}{{\rm h.c.}}
\renewcommand{\logo}{\relax}

\title{Two dark matter candidates in three-Higgs-doublet models \\ with $S_3$ symmetry}
\ShortTitle{Two dark matter candidates in three-Higgs-doublet models with $S_3$ symmetry}

\author*[a]{Anton Kun\v cinas}
\author[b]{Odd Magne Ogreid}
\author[c]{Per Osland}
\author[a]{M. Nesbitt Rebelo}

\affiliation[a]{Centro de F\'isica Te\'orica de Part\'iculas, CFTP, Departamento de F\'\i sica,\\ Instituto Superior T\'ecnico, Universidade de Lisboa,\\
Avenida Rovisco Pais nr. 1, 1049-001 Lisboa, Portugal}

\affiliation[b]{Western Norway University of Applied Sciences,\\ Postboks 7030, N-5020 Bergen, 
Norway}

\affiliation[c]{Department of Physics and Technology, University of Bergen, \\
Postboks 7803, N-5020  Bergen, Norway}

\emailAdd{Anton.Kuncinas@tecnico.ulisboa.pt}
\emailAdd{omo@hvl.no}
\emailAdd{Per.Osland@uib.no}
\emailAdd{rebelo@tecnico.ulisboa.pt}

\abstract{
Models with an extended scalar electroweak sector can have vanishing vacuum expectation values in a basis where an underlying symmetry is imposed. Such extensions are very well motivated. If a symmetry prevents couplings between fermions and additional scalars, such scalars could become viable dark matter candidates if some additional criteria are satisfied. We catalogue $S_3$-symmetric three-Higgs-doublet models, also allowing for softly broken $S_3$-symmetric scalar potential terms, based on whether a specific model could possibly accommodate a dark matter candidate. The variety of the $S_3$-symmetric family models arises due to different possibilities to arrange vacuum expectation values. Such models can have vacua with one or two vanishing vacuum expectation values. In our study we assume that the dark matter candidate is stabilised by the $\mathbb{Z}_2$ symmetry. The $\mathbb{Z}_2$ symmetry is a remnant of $S_3$ symmetry which survived spontaneous symmetry breaking, and not superimposed over $S_3$. We explore two models; with an without CP violation. These models have a single dark and two active scalar sectors. The active sectors behave in many aspects like a Type-I two-Higgs-doublet model. The dark matter candidate masses, in two cases, are different from the known (previously studied) models with three scalar doublets. After investigating the models in detail, identifying parameters compatible with both theoretical and experimental constraints, we found that the dark matter candidate mass could be within the range of $[52.5,\,89]~\text{GeV}$ or $[6.5,\,44.5]~\text{GeV}$ for a model with CP violation.
}

\FullConference{%
  7th Symposium on Prospects in the Physics of Discrete Symmetries (DISCRETE 2020-2021)\\
  29th November - 3rd December 2021\\
 Bergen, Norway}

\begin{document}

\maketitle

\section{Introduction}

Cosmological observations expose our limited knowledge of the Universe, namely that the standard cosmological model proposes that around 85\% of the matter of the Universe is made up of hypothetical Dark Matter (DM), of which we have little to no knowledge. There are many models attempting to account for the DM of the Universe. This paper summarises two studies \cite{Khater:2021wcx,Kuncinas:2022whn} based on the assumption that the scalar electroweak sector of the Standard Model (SM) could be enlarged by two SU(2) Higgs doublets and constrained by an $S_3$ symmetry~\cite{Derman:1978rx}, while accounting for DM. In the defining representation, $S_3$ is the symmetry under permutations of three objects.

One of the simplest extensions of the SM which addresses the issue of DM is the Inert Doublet Model (IDM)
~\cite{IDM}
. In the IDM, the lightest neutral member of a $\mathbb{Z}_2$-odd doublet could become a DM candidate. Models with three Higgs doublets (3HDM) have also been proposed and studied:
\begin{enumerate}
\item
Models with two non-inert doublets along with one inert doublet \cite{Grzadkowski:2009bt,Merchand:2019bod};
\item
Models with one non-inert doublet along with two inert doublets \cite{3HDM_Z2, 3HDM_CP}.
\end{enumerate}

\begin{figure}[htb]
\floatbox[{\capbeside\thisfloatsetup{capbesideposition={right,top},capbesidewidth=6.5cm}}]{figure}[\FBwidth]
{\caption{ Sketch of allowed DM mass ranges up to 1~TeV in various models. Blue: IDM\\ according to Refs.~\cite{IDM_res}
, the pale region indicates a non-saturated relic density. Red: IDM2 \cite{Merchand:2019bod}. Ochre: 3HDM without \cite{3HDM_Z2}
 and with CP violation \cite{3HDM_CP}
  Green: $S_3$-symmetric 3HDM without CP violation (R-II-1a)~\cite{Khater:2021wcx} and with CP violation (C-III-a)~\cite{Kuncinas:2022whn}.}\label{Fig:mass-ranges}}
{\includegraphics[scale=0.265]{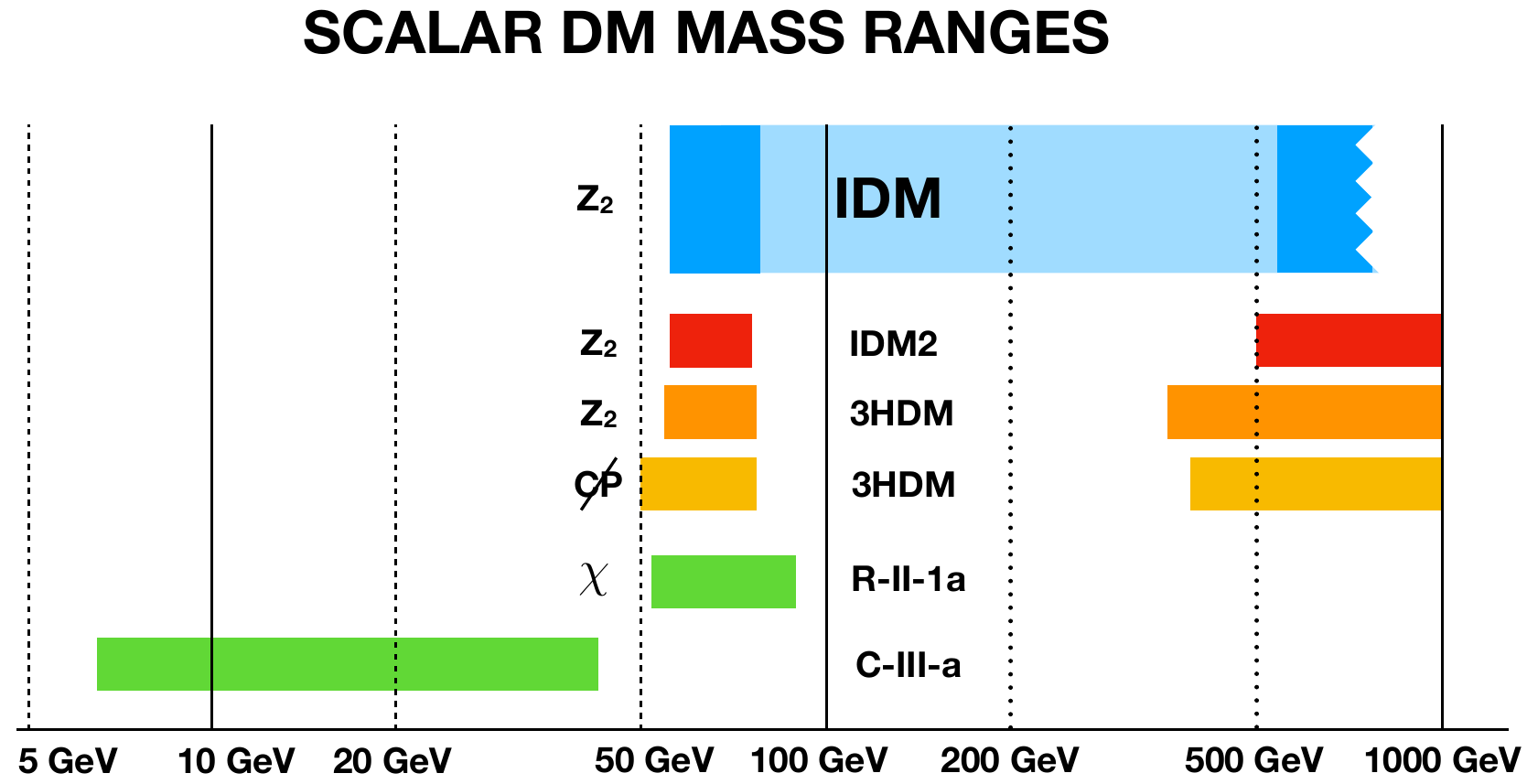}}
\end{figure}

This contribution is based on two models with an underlying $S_3$ symmetry, classified in \cite{Emmanuel-Costa:2016vej}. The models are denoted by R-II-1a~\cite{Khater:2021wcx} and C-III-a~\cite{Kuncinas:2022whn}. The C-III-a model allows for spontaneous CP violation. The DM candidate mass ranges of several models are presented in figure~\ref{Fig:mass-ranges}.

\section{The \boldmath$S_3$-symmetric models}

In terms of the $S_3$ singlet ($h_S$) and doublet $\left(h_1,\,~h_2\right)$ fields, the $S_3$-symmetric scalar potential in the irreducible representation can be written as \cite{3HDM_pot}:
\begin{subequations} \label{Eq:V-DasDey}
\begin{align}
V_2&=\mu_0^2 h_S^\dagger h_S +\mu_1^2(h_1^\dagger h_1 + h_2^\dagger h_2), \\
V_4&=
\lambda_1(h_1^\dagger h_1 + h_2^\dagger h_2)^2 
+\lambda_2(h_1^\dagger h_2 - h_2^\dagger h_1)^2
+\lambda_3[(h_1^\dagger h_1 - h_2^\dagger h_2)^2+(h_1^\dagger h_2 + h_2^\dagger h_1)^2]
\nonumber \\
&+ \lambda_4[(h_S^\dagger h_1)(h_1^\dagger h_2+h_2^\dagger h_1)
+(h_S^\dagger h_2)(h_1^\dagger h_1-h_2^\dagger h_2)+\hc] 
+\lambda_5(h_S^\dagger h_S)(h_1^\dagger h_1 + h_2^\dagger h_2) \nonumber \\
&+\lambda_6[(h_S^\dagger h_1)(h_1^\dagger h_S)+(h_S^\dagger h_2)(h_2^\dagger h_S)] 
+\lambda_7[(h_S^\dagger h_1)(h_S^\dagger h_1) + (h_S^\dagger h_2)(h_S^\dagger h_2) +\hc]
\nonumber \\
&+\lambda_8(h_S^\dagger h_S)^2.
\label{Eq:V-DasDey-quartic}
\end{align}
\end{subequations}
We have chosen to work with real coefficients. However, there remains the possibility of breaking CP spontaneously. The scalar $S_3$-symmetric  potential has an additional $\mathbb{Z}_2$ symmetry under which $h_1 \leftrightarrow - h_1$. In the irreducible representation, the $S_3$ fields can be decomposed as
\begin{equation} \label{Eq:hi_hS}
h_i=\left(
\begin{array}{c}h_i^+\\ (w_i+\eta_i+i \chi_i)/\sqrt{2}
\end{array}\right), \quad i=1,2\,\,, \qquad
h_S=\left(
\begin{array}{c}h_S^+\\ (w_S+ \eta_S+i \chi_S)/\sqrt{2}
\end{array}\right),
\end{equation}
where the $w_i$ and $w_S$ are vacuum expectation values (vev) that can be complex~\cite{Emmanuel-Costa:2016vej}.

\section{Scalar dark matter candidates}

In Ref.~\cite{Khater:2021wcx} we identified all possible $S_3$-symmetric 3HDMs which could accommodate a DM candidate. Some models have minimisation conditions requiring $\lambda_4=0$, giving rise to an additional O(2) symmetry~\cite{Emmanuel-Costa:2016vej}. Apart from this, additional continuous symmetries~\cite{Kuncinas:2020wrn} could arise, which can be spontaneously broken. These models are associated with Goldstone bosons and would require soft breaking of the $S_3$ symmetry in the potential to eliminate massless states. 

There are two approaches when constructing the Yukawa Lagrangian, depending on how the fermions transform under the $S_3$ symmetry. If they transform trivially, they can only couple to $h_S$ and there are no constraints on the couplings. Realistic masses and mixing can be generated. This is no longer possible when a DM candidate resides in the $S_3$ singlet and therefore $w_S=0$. Another possibility is when fermions transform non-trivially under $S_3$. In this case fermions are grouped into $S_3$ doublets and singlets, and it may not be possible to obtain realistic masses and mixing. 

Most $S_3$-symmetric models with a vanishing vev, which is required to stabilise DM, yield massless scalars or an unrealistic fermionic sector. Despite the variety of models presented in Ref.~\cite{Khater:2021wcx} (eleven in total), barring soft symmetry breaking, only three models survive the requirements of having a good DM candidate: no additional massless scalars present and realistic fermion masses and mixing. In these three models the fermions transform trivially under $S_3$. 

We explored two models: R-II-1a \cite{Khater:2021wcx} and C-III-a \cite{Kuncinas:2022whn} with vevs $(0,w_2,w_S)$ and $(0,\hat w_2 e^{i \sigma},\hat w_S)$, respectively. Both models share a common vacuum pattern, apart from the $\sigma$ phase in the C-III-a model. However, as discussed in Ref.~\cite{Kuncinas:2022whn}, these models do not coincide in the limit of a vanishing phase for C-III-a. The R-II-1a and C-III-a correspond to different regions of the parameter space of the $S_3$-symmetric 3HDM potential, which yields different acceptable DM candidate mass ranges. The full particle content, including interactions, is presented in Refs.~\cite{Khater:2021wcx,Kuncinas:2022whn}.

\section{Models analysis}

The two models are described in terms of eight input parameters. In R-II-1a two angles and six masses are used. In C-III-a, due to CP violation, we trade three masses for three angles to simplify computation; five angles and three masses are used as input. We scan over these parameters in order to identify regions that are compatible with a possible DM candidate. Several constraints are imposed, with each subsequent constraint being superimposed over the previous ones:
\begin{itemize}
\item Cut~1: perturbativity, stability, unitarity checks, a selection of relevant LEP constraints;
\item Cut~2: SM-like gauge and Yukawa sector, $S$ and $T$ variables, $\overline B \to X(s)\gamma$ decays;
\item Cut~3: SM-like Higgs particle decays, DM relic density, direct searches;
\end{itemize}

The applied numerical bounds are taken from PDG~\cite{Zyla:2020zbs}. We impose 3-$\sigma$ tolerance along with an additional ten per cent computational uncertainty in relevant checks. Only Cut~3 checks rely on public codes. We used $\mathsf{micrOMEGAs~5.2.7}$~\cite{Belanger:2008sj} to evaluate these constraints. 

\section{Discussion}

In the R-II-1a model due to no mixing between the inert neutral states there are potentially two possible DM candidates. However, only one of these states is a possible DM candidate, namely $\chi_1$, see eq.~\eqref{Eq:hi_hS}. The state $\eta_1$ with $m_{\eta_1}^2 \sim \lambda_4 v^2$ does not satisfy Cut~3. In contrast to R-II-1a, there is a single DM candidate in C-III-a due to mixing of the inert neutral fields. Moreover, there is mixing among all neutral active scalars in C-III-a, and hence the SM-like Higgs boson is CP indefinite.

\begin{figure}[h]
\begin{center}
\includegraphics[scale=0.25]{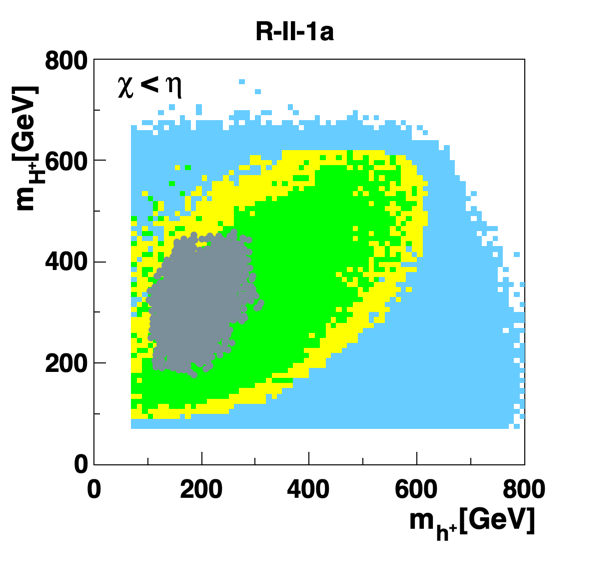}
\includegraphics[scale=0.25]{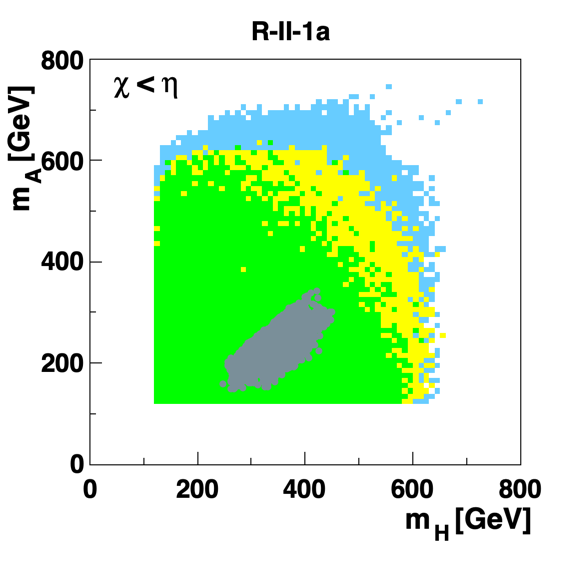}
\includegraphics[scale=0.25]{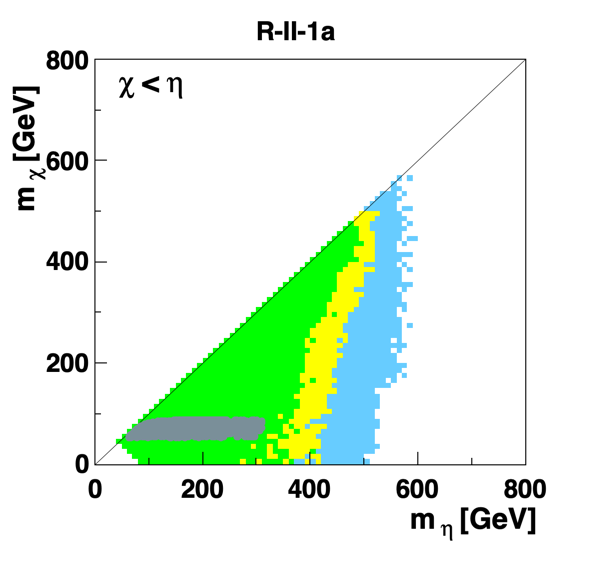}
\hspace*{-4pt}\includegraphics[scale=0.22]{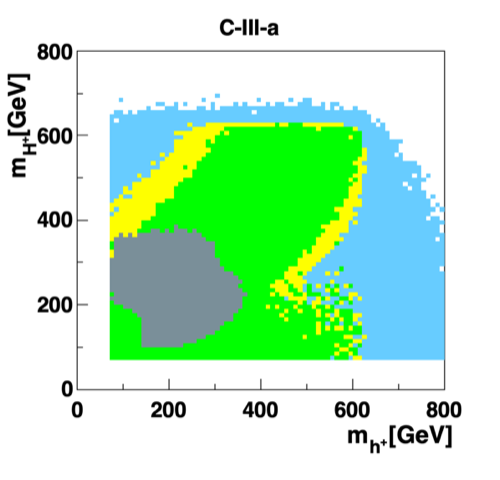}
\hspace*{3pt}\includegraphics[scale=0.25]{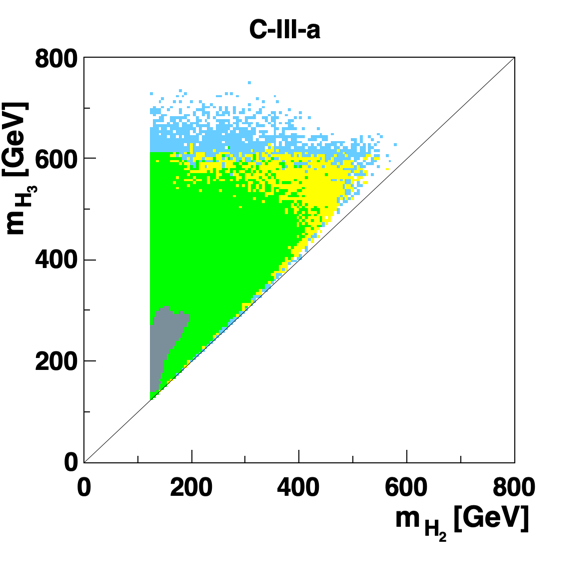}
\includegraphics[scale=0.25]{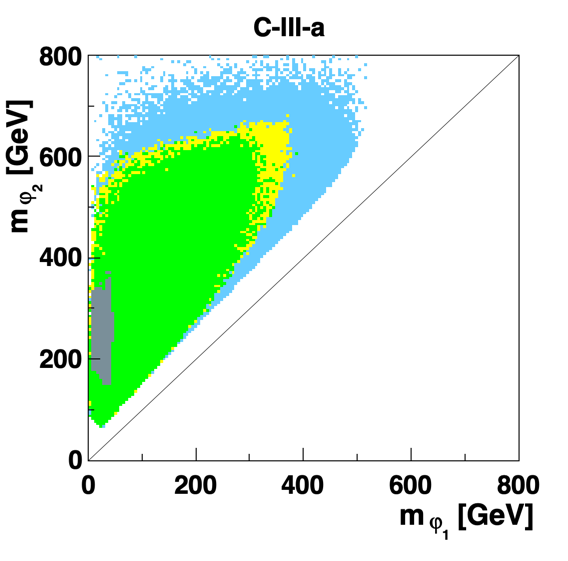}
\end{center}
\vspace*{-8mm}
\caption{Scatter plots of masses that satisfy different sets of successive Cuts. Left column: the charged sector. Middle column: the active heavy neutral sector. Right column: the inert neutral sector. The blue region satisfies Cut~1. The yellow region accommodates a 3-$\sigma$ tolerance with respect to Cut~2, whereas the green region accommodates the 2-$\sigma$ bound. The grey region is compatible with Cut~1, Cut~2 and Cut~3.}
\label{Fig:masses}
\end{figure}

In figure~\ref{Fig:masses} we present mass scatter plots after applying the previously discussed constraints. We allowed for scalars to be as heavy as 1~TeV. However, after applying Cut~1 an upper bound of around 600-800 GeV develops. The allowed regions shrink after applying Cut~2 and Cut~3. Both models favour light states of 200-400 GeV as indicated by the surviving grey regions. Such light states are not completely ruled out by the LHC searches due to the suppressed couplings and imposed constraints on the SM-like Higgs boson. While the active scalars of R-II-1a can in principle decay into other active scalars, if kinematically allowed, in C-III-a the dominant decay channel for all of the active scalars, except the SM-like Higgs, is into states with at least one DM candidate with the second scalar also coming from the inert doublet. Such processes would be accompanied by large missing transverse momentum in the detector.

After applying all three cuts over the parameter space we found that the viable DM mass regions differ drastically from the multi-doublet DM models proposed earlier, see figure~\ref{Fig:mass-ranges}. There is no high mass DM region in our models. In those the portal couplings grow fast with the DM candidate mass. High portal couplings would lead to a DM candidate annihilating too fast in the Early Universe. Heavy DM candidates in IDM-like models require the tuning of portal couplings as well as near mass degeneracy between the inert scalars. In the C-III-a model near degeneracy is impossible and we obtain a mass gap of around 70 GeV for the parameter region of interest for DM.

\begin{figure}[h]
\floatbox[{\capbeside\thisfloatsetup{capbesideposition={right,top},capbesidewidth=5cm}}]{figure}[\FBwidth]
{\caption{The spin-independent DM-nucleon cross section compatible with XENON1T~\cite{Aprile:2018dbl} data at 90\% C.L. The points represent cases that satisfy Cut~3. The red line corresponds to an approximate neutrino floor.}\label{Fig:DM_DD}}
{\includegraphics[scale=0.192]{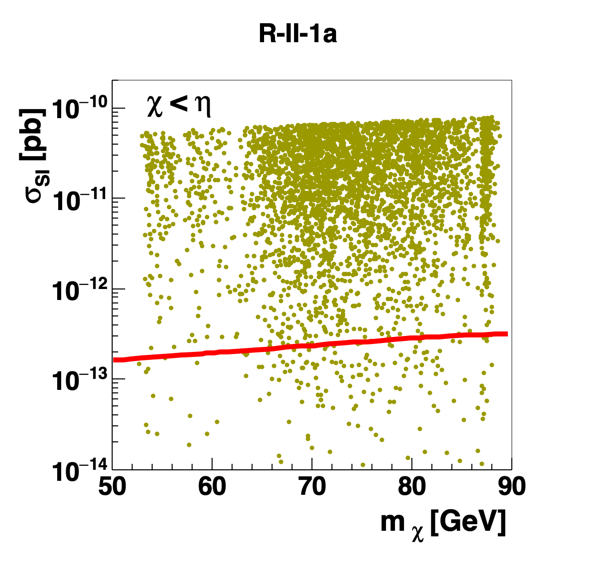}
\includegraphics[scale=0.192]{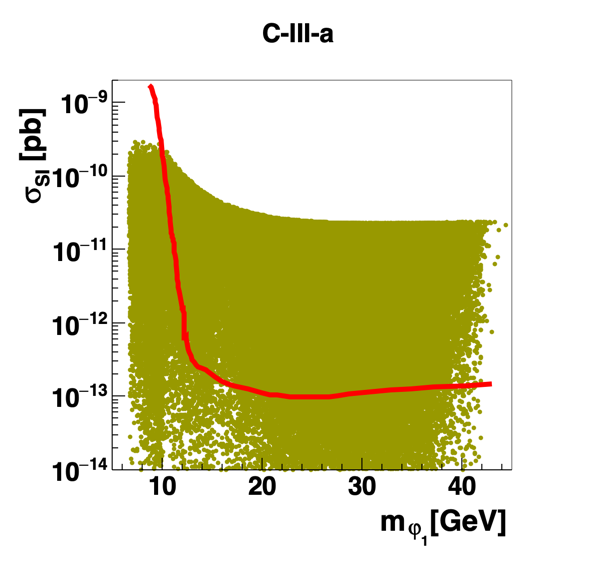}}
\end{figure}

Another interesting property, particular to C-III-a, is that there are only DM candidates with masses {$m_\mathrm{DM} < 50\text{ GeV}$}, a region that is ruled out for the other models, whereas masses above this limit are excluded due to the strength of the portal couplings. 
Furthermore, in the C-III-a model the direct DM detection criteria are satisfied for light DM states, presented in figure~\ref{Fig:DM_DD}. There are points for both models with spin-independent DM-nucleon cross section several orders of magnitudes lower than what would be probed by future DM experiments.

Deciphering the nature of DM remains one of the most important challenges in both particle physics and cosmology. After applying a selected set of constraints we determined possible DM mass ranges. Those are $[52.5,\,89]~\text{GeV}$ for R-II-1a and $[6.5,\,44.5]~\text{GeV}$ for C-III-a. These models look very promising, showing that the $S_3$-symmetric 3HDM has a very rich structure.

\acknowledgments
The work of PO~is supported in part by the Research Council of Norway, and that of AK and MNR was partially supported by Funda\c c\~ ao para a Ci\^ encia e a Tecnologia (FCT, Portugal) through the projects CFTP-FCT Unit UIDB/00777/2020 and UIDP/00777/2020, CERN/FIS-PAR/0008/2019 and PTDC/FIS-PAR/29436/2017  which are partially funded through POCTI (FEDER), COMPETE, QREN and EU. Furthermore, the work of AK has been supported by the FCT PhD fellowship with reference UI/BD/150735/2020. MNR and PO benefited from discussions that took place at the University of Warsaw during visits supported by the HARMONIA project of the National Science Centre, Poland, under contract UMO-2015/18/M/ST2/00518 (2016-2019). We also thank the University of Bergen and CFTP/IST/University of Lisbon, where collaboration visits took place. 

\bibliographystyle{JHEP}

\bibliography{ref}

\providecommand{\href}[2]{#2}\begingroup\raggedright\begin{thebibliography}{10}

\bibitem{Khater:2021wcx}
W.~Khater, A.~Kun\v{c}inas, O.~M. Ogreid, P.~Osland and M.~N. Rebelo,
  \emph{{Dark matter in three-Higgs-doublet models with S$_{3}$ symmetry}},
  \href{http://dx.doi.org/10.1007/JHEP01(2022)120}{\emph{JHEP} {\bf 01} (2022)
  120}, [\href{http://arxiv.org/abs/2108.07026}{{\tt 2108.07026}}].

\bibitem{Kuncinas:2022whn}
A.~Kun\v{c}inas, O.~M. Ogreid, P.~Osland and M.~N. Rebelo, \emph{{Dark matter
  in a CP-violating three-Higgs-doublet model with $S_3$ symmetry}},
  \href{http://arxiv.org/abs/2204.05684}{{\tt 2204.05684}}.

\bibitem{Derman:1978rx}
E.~Derman, \emph{{Flavor Unification, $\tau$ Decay and $b$ Decay Within the Six
  Quark Six Lepton {Weinberg-Salam} Model}},
  \href{http://dx.doi.org/10.1103/PhysRevD.19.317}{\emph{Phys. Rev. D} {\bf 19}
  (1979) 317--329}.

\bibitem{IDM}
N.~G. Deshpande and E.~Ma, \emph{{Pattern of Symmetry Breaking with Two Higgs
  Doublets}}, \href{http://dx.doi.org/10.1103/PhysRevD.18.2574}{\emph{Phys.
  Rev.} {\bf D18} (1978) 2574};
E.~Ma, \emph{{Verifiable radiative seesaw mechanism of neutrino mass and dark
  matter}}, \href{http://dx.doi.org/10.1103/PhysRevD.73.077301}{\emph{Phys.
  Rev.} {\bf D73} (2006) 077301},
  [\href{http://arxiv.org/abs/hep-ph/0601225}{{\tt hep-ph/0601225}}];
R.~Barbieri, L.~J. Hall and V.~S. Rychkov, \emph{{Improved naturalness with a
  heavy Higgs: An Alternative road to LHC physics}},
  \href{http://dx.doi.org/10.1103/PhysRevD.74.015007}{\emph{Phys. Rev.} {\bf
  D74} (2006) 015007}, [\href{http://arxiv.org/abs/hep-ph/0603188}{{\tt
  hep-ph/0603188}}];
Q.-H. Cao, E.~Ma and G.~Rajasekaran, \emph{{Observing the Dark Scalar Doublet
  and its Impact on the Standard-Model Higgs Boson at Colliders}},
  \href{http://dx.doi.org/10.1103/PhysRevD.76.095011}{\emph{Phys. Rev.} {\bf
  D76} (2007) 095011}, [\href{http://arxiv.org/abs/0708.2939}{{\tt
  0708.2939}}].

\bibitem{Grzadkowski:2009bt}
B.~Grzadkowski, O.~M. Ogreid and P.~Osland, \emph{{Natural Multi-Higgs Model
  with Dark Matter and CP Violation}},
  \href{http://dx.doi.org/10.1103/PhysRevD.80.055013}{\emph{Phys. Rev.} {\bf
  D80} (2009) 055013}, [\href{http://arxiv.org/abs/0904.2173}{{\tt
  0904.2173}}].

\bibitem{Merchand:2019bod}
M.~Merchand and M.~Sher, \emph{{Constraints on the Parameter Space in an Inert
  Doublet Model with two Active Doublets}},
  \href{http://dx.doi.org/10.1007/JHEP03(2020)108}{\emph{JHEP} {\bf 03} (2020)
  108}, [\href{http://arxiv.org/abs/1911.06477}{{\tt 1911.06477}}].

\bibitem{3HDM_Z2}
V.~Keus, S.~F. King, S.~Moretti and D.~Sokolowska, \emph{{Dark Matter with Two
  Inert Doublets plus One Higgs Doublet}},
  \href{http://dx.doi.org/10.1007/JHEP11(2014)016}{\emph{JHEP} {\bf 11} (2014)
  016}, [\href{http://arxiv.org/abs/1407.7859}{{\tt 1407.7859}}];
V.~Keus, S.~F. King, S.~Moretti and D.~Sokolowska, \emph{{Observable Heavy
  Higgs Dark Matter}},
  \href{http://dx.doi.org/10.1007/JHEP11(2015)003}{\emph{JHEP} {\bf 11} (2015)
  003}, [\href{http://arxiv.org/abs/1507.08433}{{\tt 1507.08433}}];
A.~Cordero, J.~Hernandez-Sanchez, V.~Keus, S.~F. King, S.~Moretti, D.~Rojas
  et~al., \emph{{Dark Matter Signals at the LHC from a 3HDM}},
  \href{http://dx.doi.org/10.1007/JHEP05(2018)030}{\emph{JHEP} {\bf 05} (2018)
  030}, [\href{http://arxiv.org/abs/1712.09598}{{\tt 1712.09598}}].

\bibitem{3HDM_CP}
A.~Cordero-Cid, J.~Hernández-Sánchez, V.~Keus, S.~F. King, S.~Moretti,
  D.~Rojas et~al., \emph{{CP violating scalar Dark Matter}},
  \href{http://dx.doi.org/10.1007/JHEP12(2016)014}{\emph{JHEP} {\bf 12} (2016)
  014}, [\href{http://arxiv.org/abs/1608.01673}{{\tt 1608.01673}}].


\bibitem{Emmanuel-Costa:2016vej}
D.~Emmanuel-Costa, O.~M. Ogreid, P.~Osland and M.~N. Rebelo, \emph{{Spontaneous
  symmetry breaking in the $S_3$-symmetric scalar sector}},
  \href{http://dx.doi.org/10.1007/JHEP08(2016)169,
  10.1007/JHEP02(2016)154}{\emph{JHEP} {\bf 02} (2016) 154},
  [\href{http://arxiv.org/abs/1601.04654}{{\tt 1601.04654}}].

\bibitem{IDM_res}
A.~Belyaev, G.~Cacciapaglia, I.~P. Ivanov, F.~Rojas-Abatte and M.~Thomas,
  \emph{{Anatomy of the Inert Two Higgs Doublet Model in the light of the LHC
  and non-LHC Dark Matter Searches}},
  \href{http://dx.doi.org/10.1103/PhysRevD.97.035011}{\emph{Phys. Rev.} {\bf
  D97} (2018) 035011}, [\href{http://arxiv.org/abs/1612.00511}{{\tt
  1612.00511}}];
J.~Kalinowski, W.~Kotlarski, T.~Robens, D.~Sokolowska and A.~F. Zarnecki,
  \emph{{Benchmarking the Inert Doublet Model for $e^+ e^-$ colliders}},
  \href{http://dx.doi.org/10.1007/JHEP12(2018)081}{\emph{JHEP} {\bf 12} (2018)
  081}, [\href{http://arxiv.org/abs/1809.07712}{{\tt 1809.07712}}].

\bibitem{3HDM_pot}
J.~Kubo, H.~Okada and F.~Sakamaki, \emph{{Higgs potential in minimal S(3)
  invariant extension of the standard model}},
  \href{http://dx.doi.org/10.1103/PhysRevD.70.036007}{\emph{Phys. Rev.} {\bf
  D70} (2004) 036007}, [\href{http://arxiv.org/abs/hep-ph/0402089}{{\tt
  hep-ph/0402089}}];
T.~Teshima, \emph{{Higgs potential in $S_3$ invariant model for quark/lepton
  mass and mixing}},
  \href{http://dx.doi.org/10.1103/PhysRevD.85.105013}{\emph{Phys. Rev.} {\bf
  D85} (2012) 105013}, [\href{http://arxiv.org/abs/1202.4528}{{\tt
  1202.4528}}];
D.~Das and U.~K. Dey, \emph{{Analysis of an extended scalar sector with $S_3$
  symmetry}}, \href{http://dx.doi.org/10.1103/PhysRevD.91.039905,
  10.1103/PhysRevD.89.095025}{\emph{Phys. Rev.} {\bf D89} (2014) 095025},
  [\href{http://arxiv.org/abs/1404.2491}{{\tt 1404.2491}}].

\bibitem{Kuncinas:2020wrn}
A.~Kuncinas, O.~M. Ogreid, P.~Osland and M.~N. Rebelo, \emph{{$S_3$-inspired
  three-Higgs-doublet models: A class with a complex vacuum}},
  \href{http://dx.doi.org/10.1103/PhysRevD.101.075052}{\emph{Phys. Rev. D} {\bf
  101} (2020) 075052}, [\href{http://arxiv.org/abs/2001.01994}{{\tt
  2001.01994}}].

\bibitem{Zyla:2020zbs}
{\scshape Particle Data Group} collaboration, P.~Zyla et~al., \emph{{Review of
  Particle Physics}}, \href{http://dx.doi.org/10.1093/ptep/ptaa104}{\emph{PTEP}
  {\bf 2020} (2020) 083C01}.

\bibitem{Belanger:2008sj}
G.~Belanger, F.~Boudjema, A.~Pukhov and A.~Semenov, \emph{{Dark matter direct
  detection rate in a generic model with micrOMEGAs 2.2}},
  \href{http://dx.doi.org/10.1016/j.cpc.2008.11.019}{\emph{Comput. Phys.
  Commun.} {\bf 180} (2009) 747--767},
  [\href{http://arxiv.org/abs/0803.2360}{{\tt 0803.2360}}].

\bibitem{Aprile:2018dbl}
{\scshape XENON} collaboration, E.~Aprile et~al., \emph{{Dark Matter Search
  Results from a One Ton-Year Exposure of XENON1T}},
  \href{http://dx.doi.org/10.1103/PhysRevLett.121.111302}{\emph{Phys. Rev.
  Lett.} {\bf 121} (2018) 111302}, [\href{http://arxiv.org/abs/1805.12562}{{\tt
  1805.12562}}].

\end{thebibliography}\endgroup

\end{document}